\documentclass[aip,jcp,reprint,noshowkeys,superscriptaddress]{revtex4-1}
\usepackage{graphicx,dcolumn,xcolor,microtype,multirow,amscd,amsmath,amssymb,amsfonts,physics,longtable,wrapfig,txfonts,siunitx}
\usepackage[version=4]{mhchem}
\usepackage{chemfig}
\usepackage{soul}

\usepackage[utf8]{inputenc}
\usepackage[T1]{fontenc}
\usepackage{txfonts}

\usepackage{siunitx}
\DeclareSIUnit[number-unit-product = {\,}]
\cal{cal}
\DeclareSIUnit\kcal{\kilo\cal}

\DeclareSIUnit\debye{D}

\usepackage[
	colorlinks=true,
    citecolor=blue,
    breaklinks=true
	]{hyperref}
\newcommand{\ie}{\textit{i.e.}}

\newcommand{\alert}[1]{\textcolor{black}{#1}}
\usepackage[normalem]{ulem}

\newcommand{\SupMat}{\textcolor{blue}{supporting information}}

\newcommand{\QP}{\textsc{quantum package}}

\newcommand{\br}{\boldsymbol{r}}
\newcommand{\bR}{\boldsymbol{R}}
\newcommand{\bp}{\boldsymbol{p}}

\newcommand{\hH}{\Hat{H}}

\newcommand{\cre}[1]{\Hat{a}_{#1}^{\dag}}
\newcommand{\ani}[1]{\Hat{a}_{#1}}

\newcommand{\eV}{(\SI{}{\eV})}

\newcommand{\MO}[1]{\phi_{#1}}

\newcommand{\cA}{\mathcal{A}}
\newcommand{\cI}{\mathcal{I}}

\newcommand{\fL}{f^\text{L}}
\newcommand{\fV}{f^\text{V}}
\newcommand{\fLV}{f^\text{LV}}
\newcommand{\bmu}{\boldsymbol{\mu}}
\newcommand{\hbmu}{\Hat{\bmu}}
\newcommand{\bla}{\boldsymbol{\lambda}}
\newcommand{\bnu}{\boldsymbol{\nu}}

\newcommand{\bgam}{\boldsymbol{\gamma}}

\newcommand{\ii}{\mathrm{i}}

\newcommand{\LCPQ}{Laboratoire de Chimie et Physique Quantiques (UMR 5626), Universit\'e de Toulouse, CNRS, UPS, France}
\newcommand{\CEISAM}{Nantes Universit\'e, CNRS,  CEISAM UMR 6230, F-44000 Nantes, France}
\newcommand{\IUF}{Institut Universitaire de France (IUF), F-75005 Paris, France}

\begin{document}

\title{Ground- and Excited-State Dipole Moments and Oscillator Strengths of Full Configuration Interaction Quality}

\author{Yann \surname{Damour}}
	\email{yann.damour@irsamc.ups-tlse.fr}
	\affiliation{\LCPQ}
\author{Ra\'ul \surname{Quintero-Monsebaiz}}
	\affiliation{\LCPQ}
\author{Michel \surname{Caffarel}}
	\affiliation{\LCPQ}
\author{Denis \surname{Jacquemin}}
	\affiliation{\CEISAM}
	\affiliation{\IUF}
\author{F\'abris \surname{Kossoski}}
	\affiliation{\LCPQ}
\author{Anthony \surname{Scemama}}
	\affiliation{\LCPQ}
\author{Pierre-Fran\c{c}ois \surname{Loos}}
	\email{loos@irsamc.ups-tlse.fr}
	\affiliation{\LCPQ}

\begin{abstract}
We report ground- and excited-state dipole moments and oscillator strengths (computed in different ``gauges'' or representations) of full configuration interaction (FCI) quality using the selected configuration interaction method known as \textit{Configuration Interaction using a Perturbative Selection made Iteratively} (CIPSI).
Thanks to a set encompassing 35 ground- and excited-state properties computed in 11 small molecules, the present near-FCI estimates allow us to assess the accuracy of high-order coupled-cluster (CC) calculations including up to quadruple excitations.
In particular, we show that incrementing the excitation degree of the CC expansion (from CCSD to CCSDT or from CCSDT to CCSDTQ) reduces the average error with respect to the near-FCI reference values by approximately one order of magnitude.
\bigskip
\begin{center}
	\boxed{\includegraphics[width=0.5\linewidth]{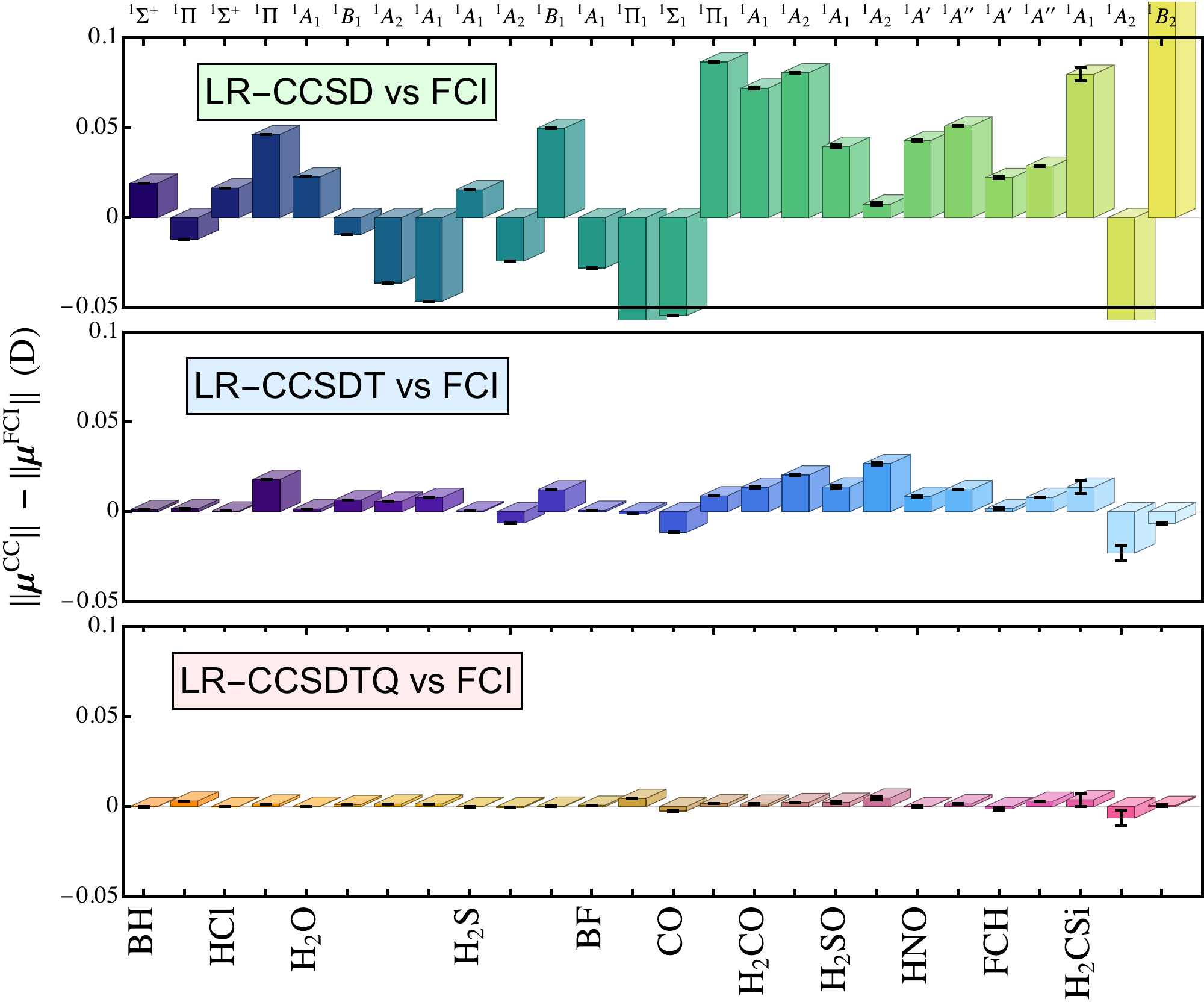}}
\end{center}
\bigskip
\end{abstract}

\maketitle

\section{Introduction}
\label{sec:intro}

The study of electric dipole moments and oscillator strengths is a major endeavor in electronic structure theory.
The electric dipole moment is a vector that characterizes the intensity and the orientation of an electric dipole, and its direction and magnitude are dictated by the distribution of the electric charges.
In a chemical system, it corresponds to the charge distribution of the electrons and nuclei and is consequently related to its electronic structure.

From an experimental point of view, the dipole moment is a physical ``signature'' of a system in a given electronic state.
Thus, it can be used to characterize unknown species or a specific isomer. \cite{Minkin_1970,Bergmann_1941,Laubengayer_1965}
In addition, the electric dipole moment is central in spectroscopy.
For example, vibrational modes are said to be infrared-active if they are accompanied by a change in the electric dipole moment. \cite{Skoog_2006}

From a more theoretical point of view, combining dipole moment and potential energy surfaces allows us to model energies and intensities of vibrational-rotational transitions, and is then useful for rovibrational spectroscopy. \cite{Nikitin_2013,Diehr_2004,Tyuterev_2017}
Furthermore, because the dipole moment is intimately linked to the charge distribution of the system in a given state, it is closely related to its electronic density and wave function.
Consequently, dipole moments are often considered descriptors of the quality of the electronic density for both ground and excited states. \cite{Hait_2018a,Giner_2021}

Another interesting physical quantity also classified as dipolar is the oscillator strength. \cite{Gupta_2016}
Because the oscillator strength is linked to the transition probability between two states (\ie, the transition dipole moment), it tells us whether or not a transition is \alert{(electric-)dipole-allowed}.
Indeed, the magnitude of the oscillator strength is directly connected to the intensity of the peaks in ultraviolet-visible spectra.

One of the main goals in theoretical quantum chemistry is to describe accurately the electronic structure of chemical systems by solving the Schr\"odinger equation, which gives access to experimentally measurable properties such as dipole moments and oscillator strengths.
Unfortunately, an accurate description of the electronic structure requires one to approach satisfactorily the solution of the Schr\"odinger equation through an appropriate and judicious set of approximations. \cite{Szabo_book,Helgaker_book,Jensen_book}

The mean-field Hartree-Fock (HF) approximation \cite{Szabo_book} is a relatively cheap method and is the starting point of correlated treatments in wave function methods.
HF is known to produce reasonably accurate properties but breaks down when correlation effects become predominant.
One textbook example is the ground-state dipole moment of \ce{CO} which is predicted with the wrong orientation at the HF level. \cite{Yoshimine_1967,Szabo_book}
This disagreement disappears when one takes into account correlation effects.
On the opposite side, the full configuration interaction (FCI) method provides the exact solution of the Schr\"odinger equation within a given one-electron basis set, by constructing the wave function as a linear combination of all possible electronic configurations. \cite{Knowles_1984,Olsen_1988,Knowles_1989,Olsen_1990,Eriksen_2021}
All these configurations, which can be represented as Slater determinants, form the so-called Hilbert space that, unfortunately, grows exponentially fast with the system size, leading to a prohibitive computational cost for real-life molecules.
Thankfully, between these two extremes, HF and FCI, a plethora of methods, some with systematic improvability, have been developed.

To reach FCI from HF, the most natural route 
 is likely to increase systematically the maximum excitation degree of the configuration interaction (CI) wave function with respect to a reference configuration (usually taken as the HF ground-state determinant).
This leads to excitation-based CI which has polynomial scaling but lacks size extensivity/consistency.
By taking into account all single and double excitations, one gets CI with singles and doubles (CISD) with a computational cost scaling as $\order*{N^6}$ (where $N$ is the number of one-electron basis functions), while adding the triples yields CI with singles, doubles, and triples (CISDT) scaling as $\order*{N^8}$, and so on.
Alternatively, one can systematically increase the seniority number (\ie, the number of unpaired electrons) or the hierarchy parameter (average of the excitation degree and half the seniority number). \cite{Ring_1980,Bytautas_2011,Kossoski_2022}
Unfortunately, all these methods require considering a huge number of electronic configurations, most of them contributing very little to the energies and/or properties of interest.

This suggests the need for a selection of determinants based on an adequate predetermined criterion to capture effectively the electronic configurations contributing the most to a given quantity.
The use of such criteria to build CI wave functions is the central idea of a general class of iterative methods known as selected CI (SCI), which sparsely explores the Hilbert space by selecting only the ``most important'' determinants for a target property. \cite{Bender_1969,Huron_1973,Buenker_1974,Evangelisti_1983,Angeli_2001c,Liu_2016b}
In most of them, this iterative selection process is performed via an energetic perturbative criterion, and determinants with the largest contributions are added to the variational space. \cite{Bender_1969,Huron_1973,Buenker_1974,Harrison_1991,Giner_2013,Holmes_2016,Schriber_2016,Tubman_2016,Sharma_2017,Tubman_2018,Coe_2018,Garniron_2019,Zhang_2020}
A second-order perturbative correction (PT2) is usually computed on top of this variational treatment. \cite{Giner_2013,Giner_2015,Holmes_2016,Sharma_2017,Garniron_2017,Garniron_2019,Zhang_2021}
The resulting SCI+PT2 methods provide a much faster energy convergence with the size of the wave function than standard CI approaches. \cite{Giner_2013,Giner_2015,Holmes_2017,Mussard_2018,Tubman_2018,Chien_2018,Tubman_2020,Loos_2020i,Yao_2020,Damour_2021,Yao_2021,Larsson_2022,Coe_2022}
Importantly, as a post-treatment, the SCI+PT2 energy and properties are usually extrapolated to the FCI limit using various strategies. \cite{Holmes_2017,Eriksen_2020a,Loos_2020i}

Relying on an exponential \emph{ansatz} of the wave function, coupled cluster (CC) methods provide an alternative, size-extensive, and systematically improvable route (with polynomial scaling) to the FCI limit. \cite{Cizek_1966,Cizek_1969,Paldus_1992,Crawford_2000,Bartlett_2007,Shavitt_2009}
Following a similar philosophy as excitation-based CI, by adding successively higher excitation levels, one gets CC with singles and doubles (CCSD),\cite{Purvis_1982,Scuseria_1987,Koch_1990a,Stanton_1993a,Stanton_1993b} CC with singles, doubles, and triples (CCSDT), \cite{Noga_1987,Scuseria_1988,Watts_1994,Kucharski_2001} CC with singles, doubles, triples, and quadruples (CCSDTQ), \cite{Kucharski_1991,Kallay_2001,Hirata_2004,Kallay_2003,Kallay_2004a} with respective computational cost scaling as $\order*{N^6}$, $\order*{N^8}$, and $\order*{N^{10}}$.
Furthermore, each of these methods can be made cheaper without altering too much their accuracy via the CC$n$ family of methods: CC2 ($N^5$), \cite{Christiansen_1995a,Hattig_2000} CC3 ($N^7$), \cite{Christiansen_1995b,Koch_1995,Koch_1997,Hald_2001,Paul_2021} and CC4 ($N^9$). \cite{Kallay_2004b,Kallay_2005,Loos_2021a,Loos_2022a}

Excited-state energies and properties can be straightforwardly obtained within the CI formalism by looking for higher roots of the CI matrix and their corresponding eigenvectors.
Likewise, one can access excited states at the CC level in the equation-of-motion (EOM) \cite{Rowe_1968a,Emrich_1981,Sekino_1984,Geertsen_1989,Stanton_1993a,Comeau_1993,Watts_1994} or linear-response (LR) \cite{Monkhorst_1977,Dalgaard_1983,Sekino_1984,Koch_1990c,Koch_1990a} frameworks.
Although they yield identical excitation energies, the excited-state properties produced by these two formalisms differ and are only equal when the FCI limit is reached. \cite{Bartlett_2007}
For the same excitation degree (hence the same computational scaling), the (non-variational) CC methods are \alert{generally} more accurate than their (variational) CI counterparts for the computation of ground- and excited-state energies and properties, \alert{especially in the Franck-Condon region}. \cite{Kallay_2003,Kallay_2004a,Kallay_2004b}
This explains why high-order CC methods have now become the workhorse of electronic structure theory when one is looking for high accuracy.
Nonetheless, their overall accuracy (with respect to FCI) remains very hard to assess, especially in the case of properties that are usually more sensitive than excitation energies to the level of theory and the one-electron basis set. \cite{Pawlowski_2004,Sarkar_2021,Giner_2018,Giner_2021,Traore_2022} 

Another feature that makes the calculation of electric (and magnetic) properties challenging is that there exist two different pathways for computing them which only become equivalent in the FCI limit \alert{but differ for approximate methods where the wave function is not fully variational with respect to all parameters}. \cite{Pople_1979,Helgaker_1989b,Jensen_book}
The first and most natural way consists in calculating the properties as expectation values of the corresponding operator associated with the physical observable of interest.
The second approach, based on the Hellmann-Feynman theorem, requires the derivative of the energy with respect to a given external perturbation linked to the observable. \cite{Pulay_1969,Pople_1979,Handy_1984}
Importantly, none of these formalisms can claim to be superior in general, \alert{although, in some cases, it has been observed that the derivative approach is likely more accurate. \cite{Diercksen_1981}}
The energy derivative technique has been first developed by Pulay in the context of self-consistent field methods, \cite{Pulay_1969} followed by others in many-body perturbation theory, \cite{Pople_1979,Fitzgerald_1985,Gauss_1987,Trucks_1988b,Gauss_1988} CI,\cite{Tachibana_1978,Goddard_1979,Krishnan_1980,Brooks_1980,Osamura_1981,Osamura_1982,Jorgensen_1983,Pulay_1983,Yamaguchi_2011} and CC methods.\cite{Adamowicz_1984,Fitzgerald_1986,Scheiner_1987,Lee_1991,Gauss_1991,Gauss_2000,Gauss_2002,Kallay_2003}
Recently, several groups have reported the implementation of nuclear gradients (\ie, energy derivatives with respect to the nuclear displacements) for \alert{SCI-SCF} \cite{Levine_2020,Park_2021,Smith_2022} and related \cite{Jiang_2022} methods.

The expectation value route is usually more straightforward in terms of implementation but one must have access explicitly to the wave function and/or to the corresponding reduced density matrices, which is not always possible.
For approximate wave functions, it has been observed that the derivative formalism is likely to lead to more accurate properties because additional contributions are taken into account. \cite{Vaval_2004,Trucks_1988}
In this context, the Lagrangian formalism, developed by Helgaker and coworkers, \cite{Helgaker_1989a,Helgaker_1989b,Koch_1990b,Jorgensen_1988} provides a rigorous mathematical framework to take into account the variation of the wave function parameters.
For example, taking or not into account the response of the orbital coefficients to the external perturbation leads to the so-called ``orbital-relaxed'' and ``orbital-unrelaxed'' properties. \cite{Hodecker_2019}
The Lagrangian formalism is employed extensively in LR-CC where the relaxation of the ground-state CC amplitudes is considered, in contrast to the cheaper EOM-CC method, resulting in size-intensive transition properties. \cite{Koch_1994}

Unfortunately, orbital relaxation effects may cause small discrepancies when employed within the frozen-core approximation since the orbital response depends on all the orbitals, even those that are frozen. \cite{Baeck_1997}
Therefore, within the frozen core approximation, the orbital-relaxed and orbital-unrelaxed dipole moments can slightly differ even at the FCI level.
This is typically the case when one considers small molecules with a significant number of frozen orbitals compared to the number of active ones.

Another degree of flexibility in the calculation of properties concerns the ``gauges'' or, more correctly, representations \cite{List_2020} (length, velocity, or mixed) chosen to compute quantities like the oscillator strength, which are only equal for the exact wave function, \ie, at the FCI limit and in a complete basis set \cite{Pedersen_1997,Pawlowski_2004} (or in the complete basis set limit for approximate methods \cite{Sauer_2019} which fulfill the Thomas-Reiche-Kuhn sum rule \cite{Thomas_1925,Reiche_1925,Kuhn_1925}).
Accordingly, gauge invariance can be employed to evaluate the degree of completeness of the one-electron basis set. \cite{Pedersen_1998}

The present work reports ground- and excited-state dipole moments as well as oscillator strengths (computed in different representations) of FCI quality obtained with the SCI method known as \textit{Configuration Interaction using a Perturbative Selection made Iteratively} (CIPSI) \cite{Huron_1973} for a set of 11 small molecules extracted from the recent work of Chrayteh \textit{et al}. \cite{Chrayteh_2021}
Thanks to the high accuracy of the present results, we can systematically assess the overall accuracy of high-order CC methods for these properties and validate the quality of the theoretical best estimates (TBEs) reported in Ref.~\onlinecite{Chrayteh_2021}.

At this stage, it is worth mentioning that works on dipole moments at the SCI level have been previously reported in the literature.
For example, the seminal work of Angeli and Cimiraglia reports a tailored selection procedure for dipole moments via a modification of the CIPSI algorithm. \cite{Angeli_2001c}
Although restricted to small wave functions, these authors achieved a significant speed-up of the convergence of the latter property and generalized it to other one-electron properties.
On the other hand, Giner \textit{et al.} studied the effect of self-consistency in the context of density-based basis-set corrections \cite{Giner_2018,Giner_2019,Loos_2019d,Giner_2020} on ground-state dipole moments using very accurate CIPSI calculations. \cite{Giner_2021}
Another study worth mentioning is the work of Eriksen and Gauss \cite{Eriksen_2020b} who reported (transition) dipole moments of \ce{LiH} and \ce{MgO} in large augmented basis sets using the many-body expanded FCI method \cite{Eriksen_2017,Eriksen_2018,Eriksen_2019a,Eriksen_2019b} which provides an interesting alternative to SCI methods. \cite{Eriksen_2020a} 
\alert{Other relevant studies have been performed using Monte Carlo CI \cite{Coe_2013} or FCI quantum Monte Carlo. \cite{Thomas_2015}}

Additionally, benchmark studies of wave function and density-based methods have been reported for both dipole moments and oscillator strengths.
For example, Hait \textit{et al.}~produced 200 benchmark values of ground-state dipole moments using CCSD(T) and basis set extrapolation to assess 88 popular or recently developed exchange-correlation functionals. \cite{Hait_2018b} 
More recently, Chrayteh \textit{et al.}~\cite{Chrayteh_2021} reported very accurate ground- and excited-state dipole moments, in addition to oscillator strengths, using LR-CC up to quintuples and applying basis set extrapolation for a set of small molecules.
In a follow-up paper, using these reference data, Sarkar \textit{et al.}~reported an extensive benchmark study of several single-reference wave function methods and time-dependent density-functional theory for several exchange-correlation functionals. \cite{Sarkar_2021}
The impact of the representations, the formalism (LR \textit{vs} EOM), and the effect of orbital relaxation (relaxed \textit{vs} unrelaxed) were carefully analyzed.
Besides these three works focussed on very accurate values for small molecules, one can also find a large panel of benchmark studies devoted to larger compounds for which it is obviously harder to establish indisputable reference values. \cite{Furche_2002,Tawada_2004,Miura_2007,Timerghazin_2008,Silva-Junior_2008,King_2008,Wong_2009,Tapavicza_2009,Guido_2010,Silva-Junior_2010d,Caricato_2010b,Hellweg_2011,Szalay_2012,Kannar_2014,Sauer_2015,Jacquemin_2016b,Jacquemin_2016c,Robinson_2018,Hodecker_2019,Hodecker_2020}

The present manuscript is organized as follows.
Section \ref{sec:theo} recalls the working equations of the CIPSI algorithm and how one computes dipole moments and oscillator strengths at the SCI level.
Section \ref{sec:comp_det} reports our computational details, while, in Sec.~\ref{sec:res}, we discuss the present results and explain in detail how we reach the FCI limit via tailored extrapolation procedures.
Our conclusions are drawn in Sec.~\ref{sec:ccl}.
Unless otherwise stated, atomic units are used throughout.

\section{Theory}
\label{sec:theo}

\subsection{Selected Configuration Interaction}
\label{sec:cipsi}

As mentioned above, SCI methods are part of the family of truncated CI methods.
Usually, their energy is defined as the sum of a variational part and a second-order perturbative contribution.
The definition of each contribution is provided below.

The (zeroth-order) variational wave function associated with the $k$th state ($k=0$ being the ground state) is
\begin{equation}
	\ket*{\Psi_k^\text{var}} = \sum_{I \in \cI} c_{Ik} \ket*{I}
\end{equation}
where $\ket*{I}$ are determinants belonging to the internal (or model) space $\cI$.
Assuming that it is normalized, this wave function has the variational energy
\begin{equation}
	E_k^\text{var} = \mel*{\Psi_k^\text{var}}{\hH}{\Psi_k^\text{var}}
\end{equation}
where $\hH$ is the usual (non-relativistic) molecular Hamiltonian
\begin{equation}
	\hH = - \sum_i \frac{\nabla_i^2}{2} - \sum_i \sum_A \frac{Z_A}{\abs*{\br_i - \bR_A}} + \sum_{i<j} \frac{1}{\abs*{\br_i - \br_j}} + \sum_{A<B} \frac{Z_A Z_B}{\abs*{\bR_A - \bR_B}}
\end{equation}
and $\br_i$ is the coordinate of the $i$th electron while $Z_A$ and  $\bR_A$ are the charge and position of the $A$th nucleus, respectively.
The associated (first-order) perturbative wave function is
\begin{equation}
	\ket*{\Psi_k^\text{pert}} = \sum_{\alpha \in \cA} c_{\alpha k} \ket*{\alpha}
\end{equation}
where the determinants $\ket*{\alpha}$, known as perturbers, belong to the external (or outer) space $\cA$.

\begin{figure}
	\includegraphics[width=\linewidth]{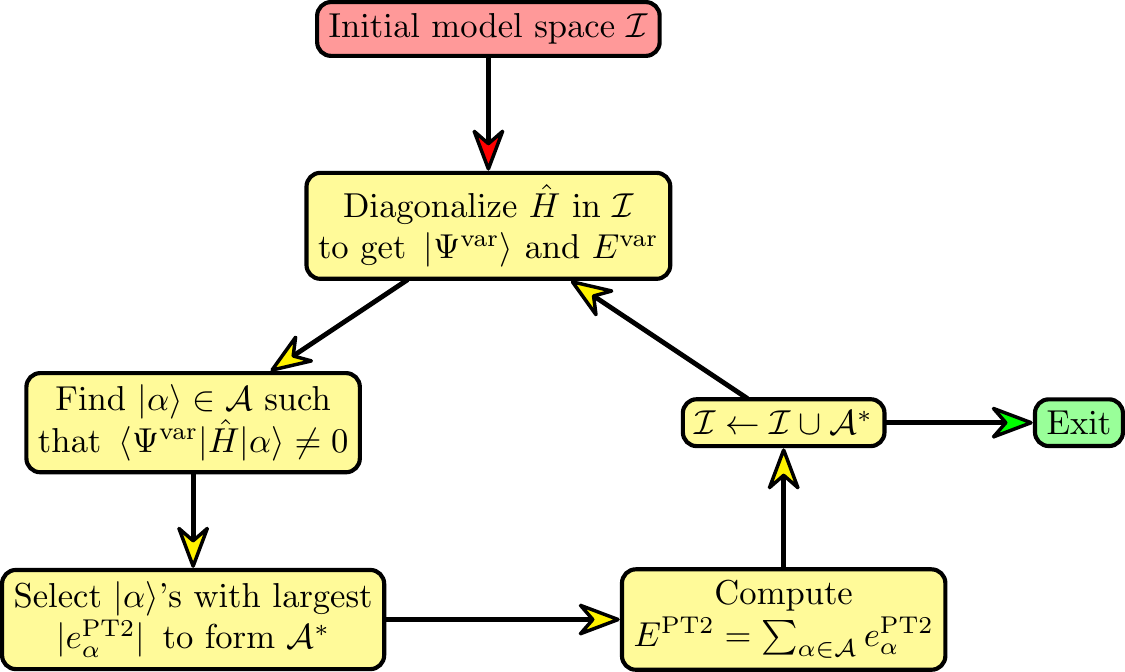}
	\caption{Iterative procedure followed by the CIPSI algorithm in the case of a single-state calculation.
	See Ref.~\onlinecite{Scemama_2019} for a description of the multi-state version.}
	\label{fig:cipsi}
\end{figure}

Employing the Epstein-Nesbet partitioning, \ie,
\begin{subequations}
\begin{align}
	\hH^{(0)} & = \sum_{IJ} \ket{I} H_{IJ} \bra{J} + \sum_{\alpha} \ket{\alpha} H_{\alpha\alpha} \bra{\alpha}
	\\
	\hH^{(1)} & = \hH - \hH^{(0)}
\end{align}
\end{subequations}
with $H_{IJ} = \mel*{I}{\hH}{J}$ and $H_{\alpha\alpha} = \mel*{\alpha}{\hH}{\alpha}$, we have $\mel*{\Psi_k^\text{var}}{\hH^{(1)}}{\Psi_k^\text{var}} = 0$ and
\begin{equation}
	E_k^\text{PT2} = \mel*{\Psi_k^\text{pert}}{\hH^{(1)}}{\Psi_k^\text{var}}
\end{equation}
where the second-order perturbative energy can be conveniently recast as
\begin{equation}
	E_k^\text{PT2} =  \sum_\alpha e_{\alpha k}^\text{PT2} = \sum_\alpha \frac{\mel*{\alpha}{\hH}{\Psi_k^\text{var}}^2}{E_k^\text{var} - H_{\alpha\alpha}}
\end{equation}
The SCI+PT2 energy of the $k$th excited state is thus given by the sum $E_k^\text{var} + E_k^\text{PT2}$.
The iterative procedure of the CIPSI algorithm is schematically represented in Fig.~\ref{fig:cipsi} in the case of a single-state calculation.
We refer the interested reader to Ref.~\onlinecite{Garniron_2019} for additional details.

\subsection{Properties as expectation values}
\label{sec:expval}
Here we follow the approach based on the expectation value of the corresponding operator to compute properties at the SCI level.

In the case of a globally neutral system, the dipole operator is
\begin{equation}
\label{eq:mu}
	\hbmu = - \sum_i \br_i + \sum_A Z_A \bR_A
\end{equation}
and the dipole moment computed from the zeroth-order wave function associated with the $k$th state is consequently
\begin{equation}
\label{eq:bmu}
	\bmu_k = \mel*{\Psi_k^\text{var}}{\hbmu}{\Psi_k^\text{var}} = - \sum_i \mel*{\Psi_k^\text{var}}{\br_i}{\Psi_k^\text{var}} + \sum_A Z_A \bR_A
\end{equation}
while the oscillator strength computed in the so-called length representation is given by
\begin{equation}
	\fL_k = \frac{2 \Delta E_k^\text{var}}{3} \bla_k \cdot \bla_k
\end{equation}
where
\begin{equation}
\label{eq:bla}
	\bla_k = - \sum_i \mel*{\Psi_0^\text{var}}{\br_i}{\Psi_k^\text{var}}
\end{equation}
is the transition dipole moment and $\Delta E_k^\text{var} = E_k^\text{var} - E_0^\text{var}$ is the vertical excitation energy associated with the $k$th excited state.
It is also possible to compute the oscillator strength in the velocity representation.
In this case, it reads
\begin{equation}
	\fV_k = \frac{2}{3\Delta E_k^\text{var} } \bnu_k \cdot \bnu_k
\end{equation}
where
\begin{equation}
\label{eq:bnu}
	\bnu_k = - \sum_i \mel*{\Psi_0^\text{var}}{\bp_i}{\Psi_k^\text{var}}
\end{equation}
and $\bp_i = - \ii \nabla_i$ is the momentum operator of electron $i$.
It is also useful to compute the mixed length-velocity representation
\begin{equation}
	\fLV_k = - \frac{2\ii}{3} \bmu_k \cdot \bnu_k
\end{equation}
which does not involve the energy difference between the two electronic states.
The quantities $\bmu_k$, $\bla_k$, and $\bnu_k$ defined in Eqs.~\eqref{eq:bmu}, \eqref{eq:bla}, and \eqref{eq:bnu} are easily computed using the Slater-Condon rules. \cite{Szabo_book,Scemama_2013a}

For practical purposes, it is convenient to recast Eq.~\eqref{eq:bmu} as
\begin{equation}
	\bmu_k = - \sum_{pq} \gamma^k_{pq} \mel*{\MO{p}}{\br}{\MO{q}} + \sum_A Z_A \bR_A
\end{equation}
where
\begin{equation}
	\gamma_{pq}^k = \mel*{\Psi_k^\text{var}}{\cre{q} \ani{p}}{\Psi_k^\text{var}}
\end{equation}
are the elements of the one-electron density matrix associated with the $k$th state, and $\cre{p}$ ($\ani{p}$) is the second quantization creation (annihilation) operator that creates (annihilates) an electron in the spatial orbital $\MO{p}(\br)$.
Similarly, for the oscillator strengths, $\bnu_k$ and $\bla_k$ can be computed with the one-electron transition density matrix, $\bgam^{0k}$, as follows
\begin{subequations}
\begin{align}
	\bnu_k & = - \sum_{pq} \gamma_{pq}^{0k} \mel*{\phi_p}{\br}{\phi_q}
	\\
	\bla_k & = - \sum_{pq} \gamma_{pq}^{0k} \mel*{\phi_p}{\bp}{\phi_q}
\end{align}
\end{subequations}
with
\begin{equation}
	\gamma_{pq}^{0k} = \mel*{\Psi_{0}^\text{var}}{a_q^\dagger a_p}{\Psi_k^\text{var}}
\end{equation}

\section{Computational Details}
\label{sec:comp_det}

\begin{figure}
	\includegraphics[width=\linewidth]{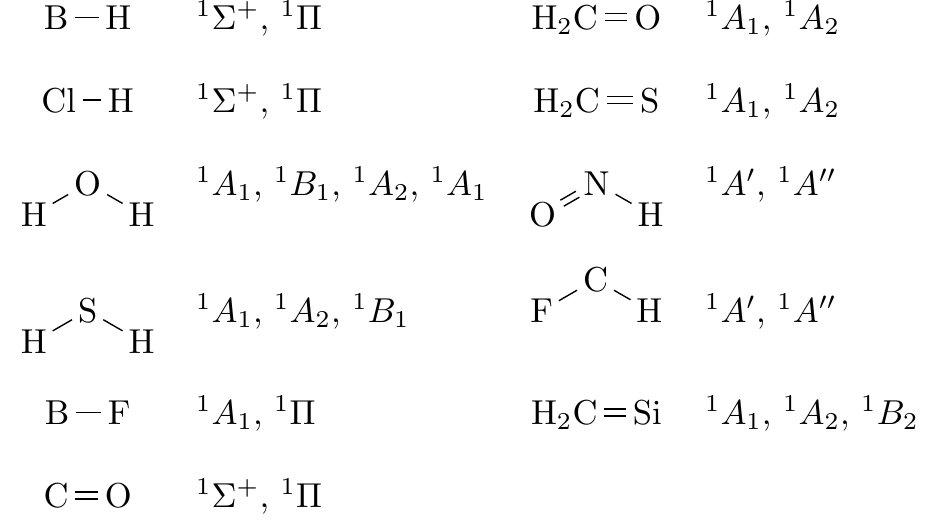}
	\caption{List of molecules and states studied in the present study.}
	\label{fig:mol}
\end{figure}

The molecules and states considered in this paper are represented in Fig.~\ref{fig:mol}.
The geometries (computed at the CC3/aug-cc-pVTZ level) and the CC results reported here have been taken from the work of Chrayteh \textit{et al.} \cite{Chrayteh_2021}
All these calculations have been performed within the frozen (large for third-row atoms) core approximation with the MRCC software. \cite{Kallay_2020}
For the sake of completeness, these geometries as well as the corresponding HF energies in the different basis sets are reported in {\SupMat}.
\alert{Here, we only consider singlet-singlet transitions, but the same procedure can be employed for higher spin states.}

Concerning the properties, the CC dipole moments have been computed within the LR formalism and are the so-called ``orbital-relaxed'' ones, which are known to be more accurate as the orbital response is properly taken into account.
The SCI oscillator strengths have been computed in the length, velocity, and mixed representations, while their CC counterparts are only available in the length representation.
All the SCI+PT2 calculations have been performed with {\QP}, \cite{Garniron_2019} where the CIPSI algorithm (see Sec.~\ref{sec:cipsi}) is implemented and where we have implemented the calculation of dipole moments and oscillator strengths at the SCI level using the expectation value formalism presented in Sec.~\ref{sec:expval}.
The raw data associated with each figure and table can be found in {\SupMat}.

For each system, starting from the HF orbitals, a first multi-state SCI calculation is performed to generate wave functions with at least \num{5e6} determinants, or large enough to reach a PT2 energy smaller than \SI{1e-6}{\hartree}.
These wave functions are then used to generate state-averaged natural orbitals.
For the smallest molecules (\ce{BH}, \ce{HCl}, \ce{H2O}, \ce{H2S}, and \ce{BF}), state-averaged optimized orbitals have been computed starting from these state-averaged natural orbitals via minimization of the variational energy at each CIPSI iteration until reaching at least \num{2e5} determinants or an energy gain between two successive iterations smaller than \SI{1e-6}{\hartree}.
More details about the orbital optimization in SCI can be found in Refs.~\onlinecite{Damour_2021} and \onlinecite{Yao_2021}.
For the remaining larger systems, we did not see any improvement going from natural to optimized orbitals.
Consequently, the calculations on the second set of molecules have been performed using the state-averaged natural orbitals.
Our goal is to reach a variational space with at least \num{5e7} determinants or large enough to reach a PT2 energy smaller than \SI{1e-6}{\hartree}.
The energies, dipole moments, and oscillator strengths are computed at each CIPSI iteration using the variational wave function and are extrapolated to the FCI limit, \ie, $E_k^\text{PT2} \to 0$, by fitting a second-degree polynomial using the last 4 points, \ie, corresponding to the four largest variational wave functions (see Sec.~\ref{sec:extrap} for additional details about the extrapolation procedure).
We refer to these results as extrapolated FCI (exFCI) values in the following.
Note that excitation energies are computed as differences of extrapolated (total) energies. \cite{Holmes_2017,Chien_2018,Loos_2018a,Loos_2019,Loos_2020d,Loos_2020c,Loos_2020f,Veril_2021}

In the statistical analysis presented below, we report the usual indicators: the mean signed error (MSE), the mean absolute error (MAE), the root-mean-square error (RMSE), the standard deviation of the errors (SDE) as well as the largest positive and negative deviations [Max($+$) and Max($-$), respectively].

\section{Results and discussion}
\label{sec:res}

\begin{table*}
\caption{Dipole moment of 26 electronic states and oscillator strengths in the length (L), velocity (V), and mixed (LV) representations for the 9 dipole-allowed transitions computed at the exFCI/aug-cc-pVTZ level.
The TBEs extracted from the work of Chrayteh \textit{et al}. \cite{Chrayteh_2021} and computed in the same basis are also listed.
An estimate of the extrapolation error associated with each value is reported in parentheses.
V and R stand for valence and Rydberg excited states, respectively.}
\label{tab:exFCI-vs-TBE}
\begin{ruledtabular}
\begin{tabular}{llccccccc}
& &
& \multicolumn{4}{c}{exFCI}
& \multicolumn{2}{c}{TBE} \\
\cline{4-7} \cline{8-9}
Molecule
& Excitation
& Nature
& $\norm{\bmu}$ & $f^\text{L}$ & $f^\text{V}$ & $f^\text{LV}$ & $\norm{\bmu}$ & $f^\text{L}$ \\
\hline
\ce{BH}       &   $^1\Sigma^+$                   &   &    1.408(0) &             &           &            & 1.409            \\
              &   $^1\Pi$                        & V &    0.554(0) &  0.048(0)   & 0.057(0)  & 0.052(0)   & 0.559 & 0.048    \\
\ce{HCl}      &   $^1\Sigma^+$                   &   &    1.084(0) &             &           &            & 1.084            \\
              &   $^1\Pi$                        & V &    2.501(0) &  0.055(0)   & 0.054(0)  & 0.054(0)   & 2.501 & 0.055    \\
\ce{H2O}      &   $^1A_1$                        &   &    1.840(0) &             &           &            & 1.840            \\
              &   $^1B_1$($n \to 3s$)            & R &    1.558(0) &  0.054(0)   & 0.056(0)  & 0.055(0)   & 1.558 & 0.054    \\
              &   $^1A_2$($n \to 3p$)            & R &    1.105(1) &             &           &            & 1.106            \\
              &   $^1A_1$($n \to 3s$)            & R &    1.214(1) &  0.100(0)   & 0.102(0)  & 0.101(0)   & 1.213 & 0.100    \\
\ce{H2S}      &   $^1A_1$                        &   &    0.977(0) &             &           &            & 0.977            \\
              &   $^1A_2$($n \to 4p$)            & R &    0.499(1) &             &           &            & 0.498            \\
              &   $^1B_1$($n \to 4s$)            & R &    1.866(1) &  0.063(0)   & 0.063(0)  & 0.063(0)   & 1.865 & 0.063    \\
\ce{BF}       &   $^1A_1$                        &   &    0.824(1) &             &           &            & 0.824            \\
              &   $^1\Pi_1$($\sigma \to \pi^*$)  & V &    0.294(1) &  0.468(0)   & 0.490(1)  & 0.479(0)   & 0.299 & 0.468    \\
\ce{CO}       &   $^1\Sigma_1$                   &   &    0.116(1) &             &           &            & 0.115            \\
              &   $^1\Pi_1$($n \to \pi^*$)       & V &    0.130(0) &  0.166(0)   & 0.173(0)  & 0.170(1)   & 0.126 & 0.166    \\
\ce{H2CO}     &   $^1A_1$                        &   &    2.384(5) &             &           &            & 2.375            \\
              &   $^1A_2$($n \to \pi^*$)         & V &    1.325(2) &             &           &            & 1.325            \\
\ce{H2CS}     &   $^1A_1$                        &   &    1.695(3) &             &           &            & 1.694            \\
              &   $^1A_2$($n \to \pi^*$)         & V &    0.839(6) &             &           &            & 0.840            \\
\ce{HNO}      &   $^1A'$                         &   &    1.676(1) &             &           &            & 1.674            \\
              &   $^1A''$($n \to \pi^*$)         & V &    1.675(3) &             &           &            & 1.676            \\
\ce{FCH}      &   $^1A'$                         &   &    1.439(2) &             &           &            & 1.438            \\
              &   $^1A''$                        &  V &    0.958(5) &  0.006(0)   & 0.008(0)  & 0.007(0)   & 0.964 & 0.006    \\
\ce{H2CSi}    &   $^1A_1$                        &   &    0.137(3) &             &           &            & 0.142            \\
              &   $^1A_2$                        & R &    1.933(2) &             &           &            & 1.924            \\
              &   $^1B_2$                        & R &    0.042(1) &  0.034(1)   & 0.032(0)  & 0.033(0)   & 0.039 & 0.034
\end{tabular}
\end{ruledtabular}
\end{table*}

The dipole moments of the 26 states investigated in the present study alongside the oscillator strengths (in the length, velocity, and mixed representations) of the 9 dipole-allowed electronic transitions are listed in Table \ref{tab:exFCI-vs-TBE}.
We also report in parentheses an estimate of the extrapolation error associated with each value (see below).
The TBEs taken from the work of Chrayteh \textit{et al}. \cite{Chrayteh_2021} are listed as well.

\subsection{Extrapolation procedure}
\label{sec:extrap}

\begin{figure*}
	\includegraphics[width=0.47\linewidth]{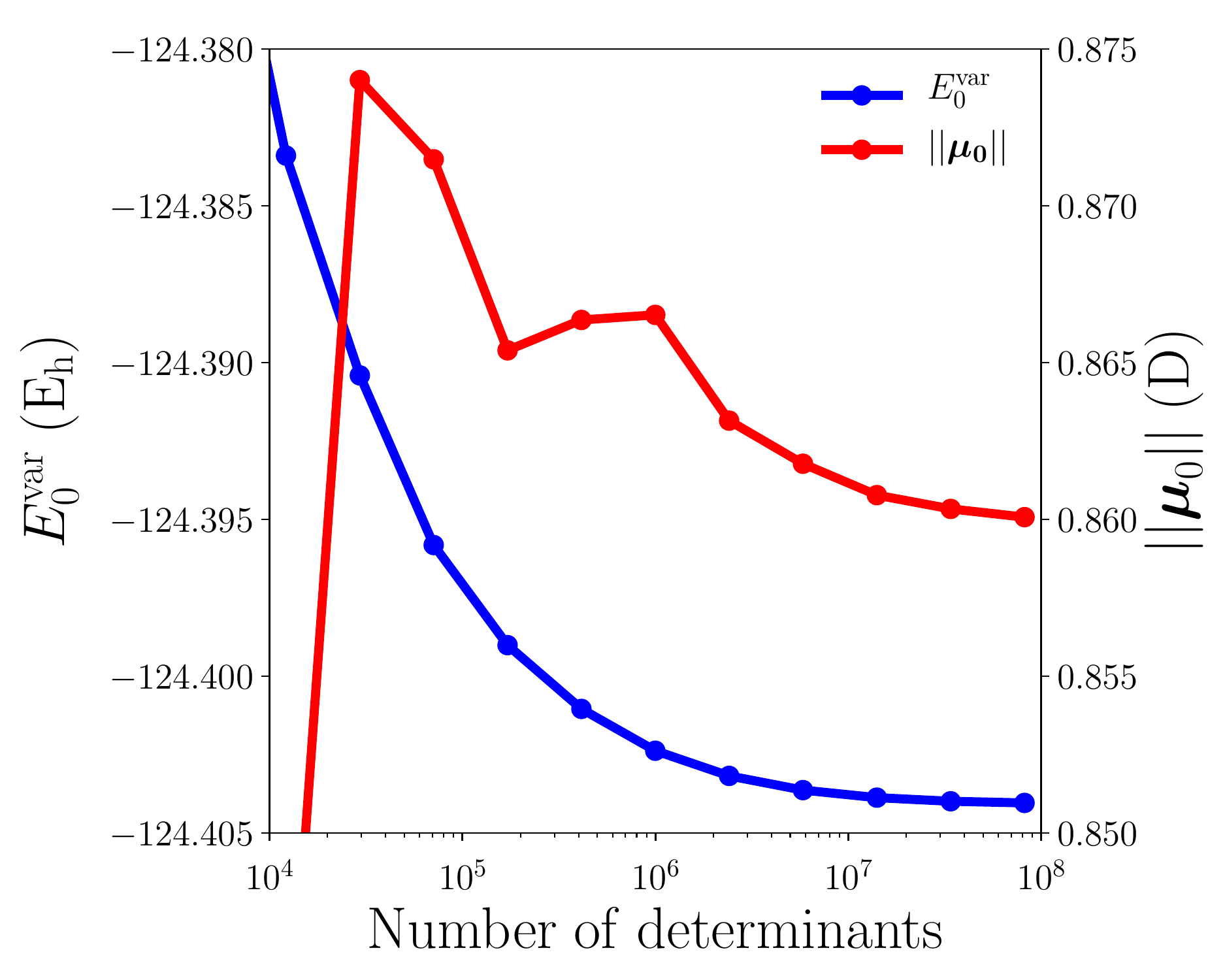}
	\hspace{0.05\linewidth}
	\includegraphics[width=0.47\linewidth]{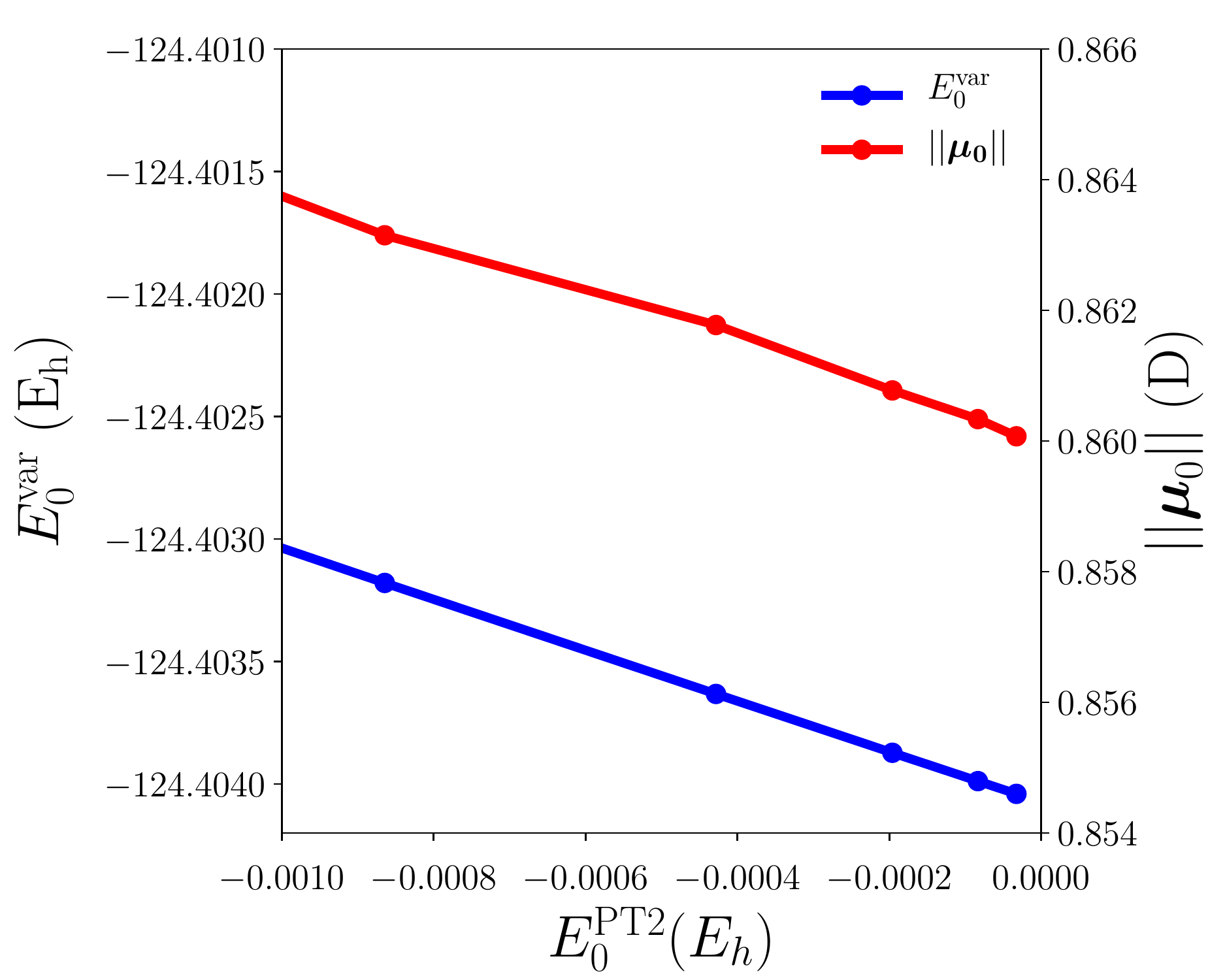}
	\caption{Ground-state variational energy $E_0^\text{var}$ of \ce{BF} (obtained with the aug-cc-pVDZ basis) and its corresponding dipole moment $\norm*{\bmu_0}$ as functions of the number of determinants in the variational wave function (left) and the second-order perturbative energy $E_0^\text{PT2}$ (right).}
	\label{fig:e_f_ndet_e_f_pt2}
\end{figure*}

As discussed above, in the CIPSI method, the wave function is built iteratively.
At each iteration, the determinants with the largest contributions to the second-order perturbative energy, $\abs*{e_{\alpha}^\text{PT2}}$, are added to the variational space (see Fig.~\ref{fig:cipsi}).
In practice, we double the size of the variational space at each iteration and include the additional determinants required to obtain eigenstates of the $\Hat{S}^2$ spin operator. \cite{Chilkuri_2021}
As a consequence of this growth, the variational energy decreases as the number of iterations increases.
This is, of course, not strictly true for properties that are not directly linked to the variational principle.
However, even if there is no direct relationship between the quality of the variational energy and a given property, the important determinants for the description of this property will eventually enter the variational space as it grows.
Consequently, although it is possible to directly select determinants for a given property as shown by Angeli and coworkers, \cite{Angeli_2001} the determinant selection based on an energy criterion is, in practice, a reasonable and universal way of producing accurate properties at the SCI level.

To illustrate these points, we report in the left panel of Fig.~\ref{fig:e_f_ndet_e_f_pt2} the evolution of the ground-state variational energy $E_0^\text{var}$ and the norm of the ground-state dipole moment $\norm*{\bmu_0}$ as functions of the number of determinants in the variational space for the \ce{BF} molecule computed in the aug-cc-pVDZ basis.
As one can see, while $E_0^\text{var}$ decreases monotonically towards the FCI limit (blue curve), the convergence of $\norm*{\bmu_0}$ (red curve) is more erratic but $\norm*{\bmu_0}$ eventually stabilizes for large enough wave functions and converges smoothly to its FCI limiting value.

To have a closer look at the region where one performs the extrapolation, we have plotted in the right panel of Fig.~\ref{fig:e_f_ndet_e_f_pt2} the evolution of the same quantities (for the same system) as functions of the second-order perturbative energy $E_0^{(2)}$.
As empirically observed, the behavior of $E_0^\text{var}$ for small $E_0^\text{PT2}$ is linear as expected from basic perturbative arguments (see blue curve in Fig.~\ref{fig:e_f_ndet_e_f_pt2}).
One can therefore safely extrapolate $E_0^\text{var}$ to $E_0^\text{PT2} = 0$ using the largest variational wave functions (or equivalently the smallest $E_0^\text{PT2}$ values) using a first- or second-order polynomial in $E_0^\text{PT2}$ to estimate the FCI energy.
A similar observation holds for the dipole moment (red curve) but the corresponding curve shows a significant quadratic character and the asymptotic regime usually appears for larger wave functions (see below).
Nonetheless, we employ the same procedure as for the energy and estimate the FCI value of the dipole moment using a quadratic fit in $E_0^\text{PT2}$ based on the four largest variational wave functions.
A rough error estimate is provided by the largest difference in extrapolated values between this 4-point fit and its 3- and 5-point counterparts.

This procedure is performed independently for each electronic state in the case of the energy and the dipole moment.
For the oscillator strength that is naturally related to the ground and the target excited state, the extrapolation procedure involves a second-order polynomial in the averaged second-order perturbative energies $(E_0^\text{PT2} + E_k^\text{PT2})/2$.

\begin{figure*}
	\includegraphics[width=0.47\linewidth]{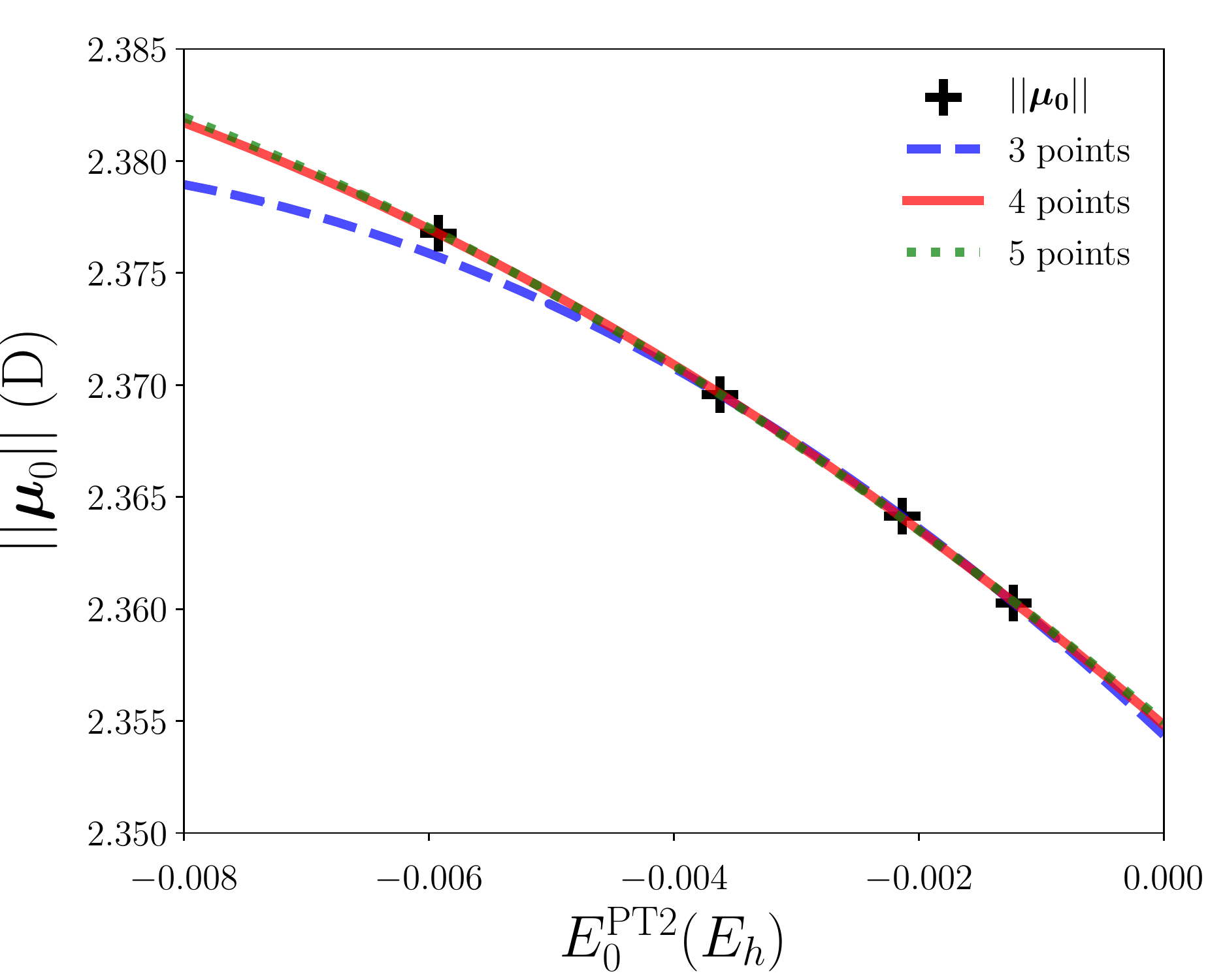}
	\hspace{0.05\linewidth}
	\includegraphics[width=0.47\linewidth]{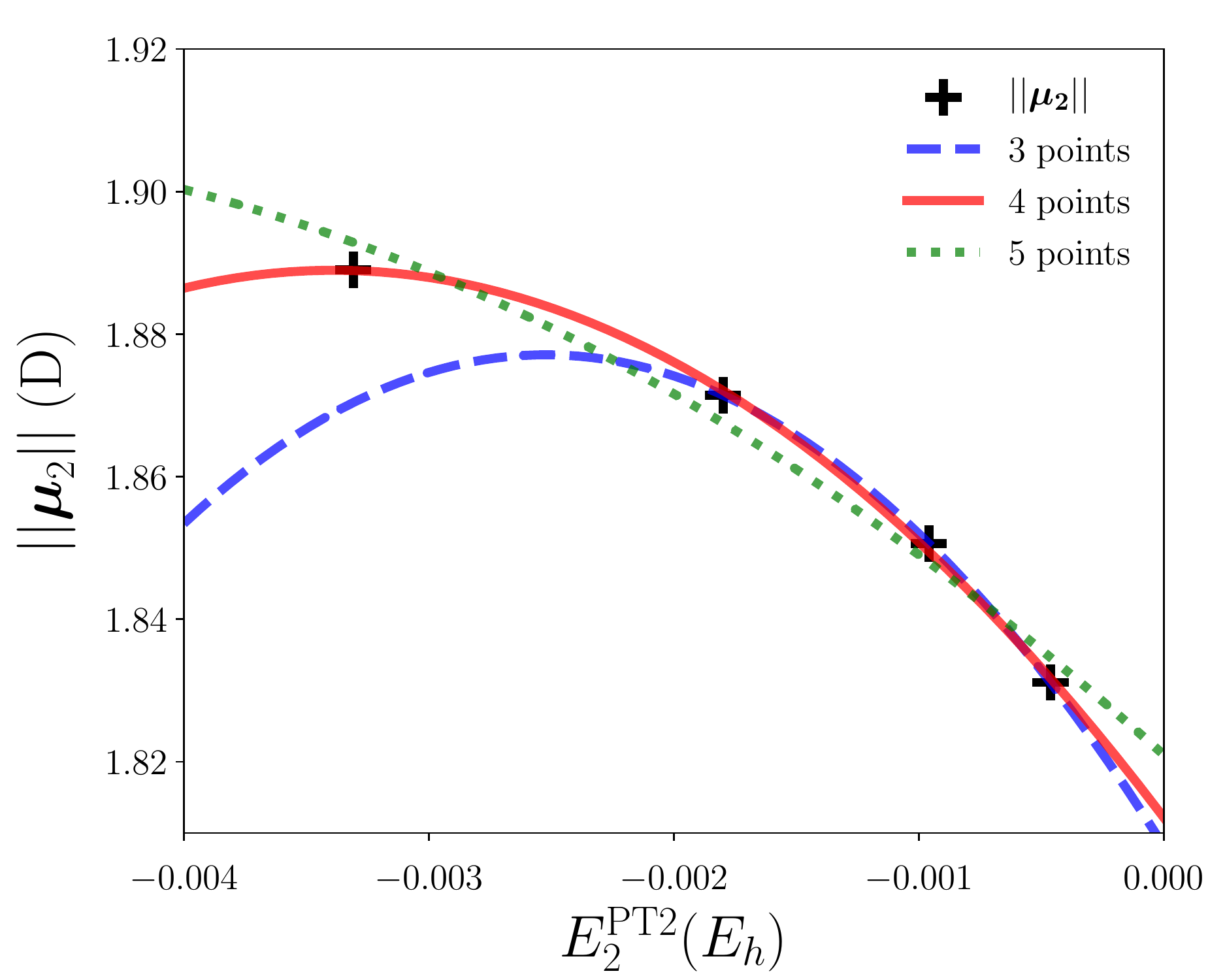}
	\caption{Left: Ground-state dipole moment of \ce{H2C=O} (obtained with the aug-cc-pVDZ basis), $\norm*{\bmu_0}$, as a function of the second-order perturbative energy $E_0^\text{PT2}$.
	Right: Second excited-state dipole moment of \ce{H2S} (obtained with the aug-cc-pVQZ basis), $\norm*{\bmu_2}$, as a function of the second-order energy $E_2^\text{PT2}$.
	The corresponding quadratic fits obtained with 3, 4, and 5 points are also reported.
	The raw data associated with these extrapolations can be found in Table \ref{tab:h2co_h2s_extr}.}
	\label{fig:h2co_h2s_extr}
\end{figure*}

\begin{table}
	\caption{Dipole moments of the ground-state of \ce{H2C=O} obtained at the exFCI/aug-cc-pVDZ level and the second excited state of \ce{H2S} obtained at the exFCI/aug-cc-pVQZ level as functions of the number of points included in the extrapolation procedure.
$\Delta$ is the deviation to the four-point extrapolation.}
	\label{tab:h2co_h2s_extr}
    \begin{ruledtabular}
        \begin{tabular}{ccccc}
            Number of & \multicolumn{2}{c}{\ce{H2C=O}} & \multicolumn{2}{c}{\ce{H2S}} \\
            \cline{2-3} \cline{4-5}
            points & $\norm*{\bmu_0}$ (\si{\debye}) & $\Delta$ (\si{\debye}) & $\norm*{\bmu_2}$ (\si{\debye}) & $\Delta$ (\si{\debye})\\
            \hline
            \rule{0pt}{12pt}
            3 & 2.3544 & 0.0004 & 1.8082 & 0.0040 \\
            4 & 2.3548 &        & 1.8122 & \\
            5 & 2.3549 & 0.0001 & 1.8210 & 0.0088 \\
        \end{tabular}
    \end{ruledtabular}
\end{table}

Illustrative examples for dipole moments are reported in Fig.~\ref{fig:h2co_h2s_extr} and the corresponding numerical values are gathered in Table \ref{tab:h2co_h2s_extr}.
The left panel of Fig.~\ref{fig:h2co_h2s_extr} shows a well-behaved case where the data are fitted quite well by a quadratic polynomial and the extrapolated value is fairly independent of the number of points.
The right panel shows an ill-behaved case where our procedure can hardly model the evolution of the dipole moment and the error is of the order of \SI{0.01}{\debye}.
Problematic cases are hard to detect \emph{a priori} and depend on the selected system, state, and basis set.

\begin{figure*}
    \includegraphics[width=0.47\linewidth]{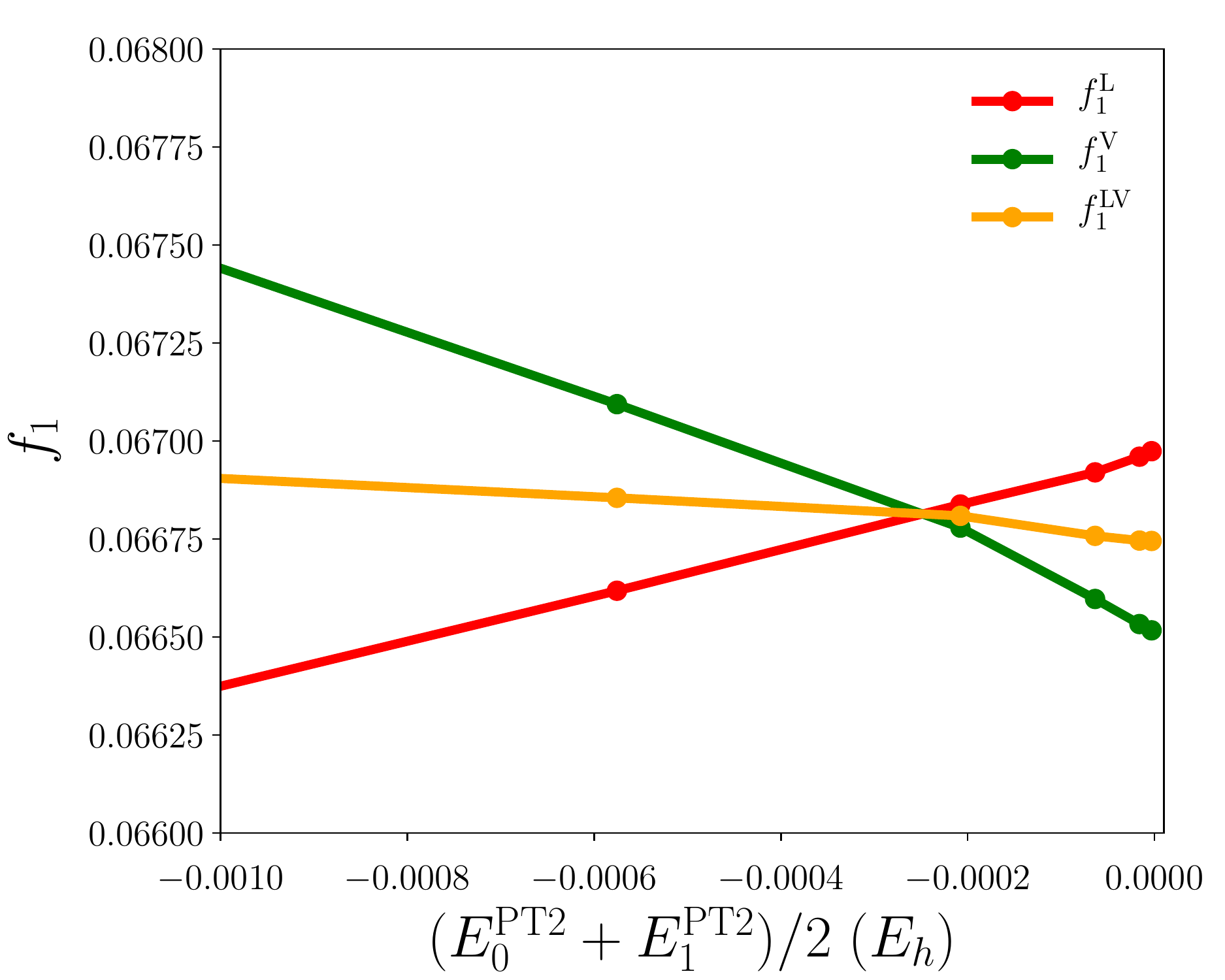}
    \hspace{0.05\linewidth}
    \includegraphics[width=0.47\linewidth]{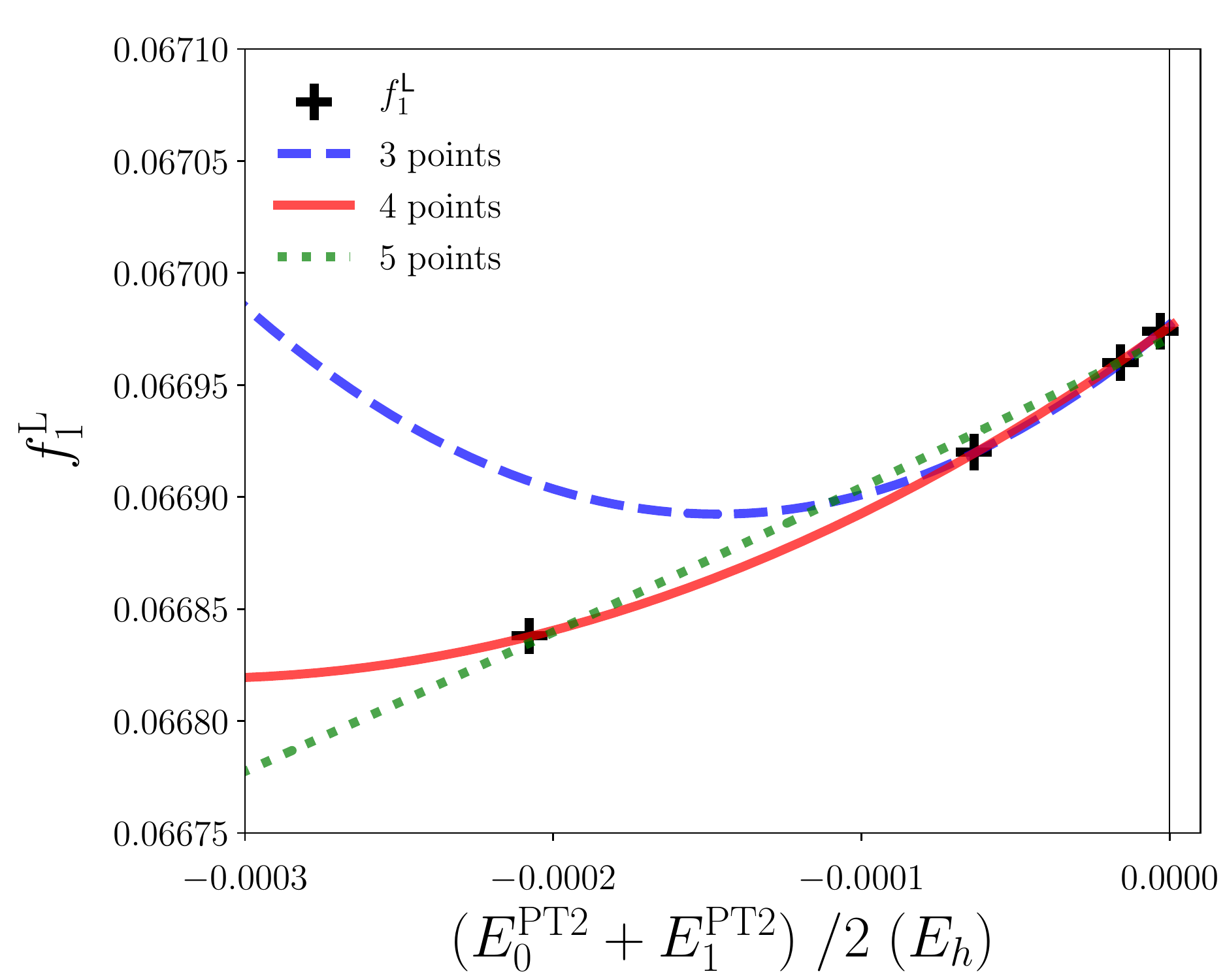}
    \caption{Oscillator strength of \ce{H2S} for the transition between the ground and lowest excited state (obtained with the aug-cc-pVDZ basis), $f_1$, in the length (L), velocity (V), and mixed (LV) representations, as a function of the averaged second-order perturbative energies $(E_0^\text{PT2}+E_1^\text{PT2})/2$.
	For the length representation, a zoom of the region where the extrapolation is performed is shown in the right panel.
    The corresponding quadratic fits obtained with 3, 4, and 5 points are also reported.}
    \label{fig:f_extr}
\end{figure*}

Figure \ref{fig:f_extr} reports the oscillator strength between the ground and first excited states of \ce{H2S} computed with the aug-cc-pVDZ basis set, in the length, velocity, and mixed representations.
In the case of oscillator strengths, we also rely on quadratic fits to estimate the FCI limiting values and the corresponding fitting errors.
The different limiting values reached with the three representations are clearly visible on the left panel.
We underline that these differences remain fairly small (below $10^{-3}$ in this particular case).
The right panel shows the extrapolation (with different numbers of points) of $f_1^\text{L}$ as a function of $(E_0^\text{PT2}+E_1^\text{PT2})/2$.

\subsection{Dipole moments}

\begin{figure*}
    \includegraphics[width=\linewidth]{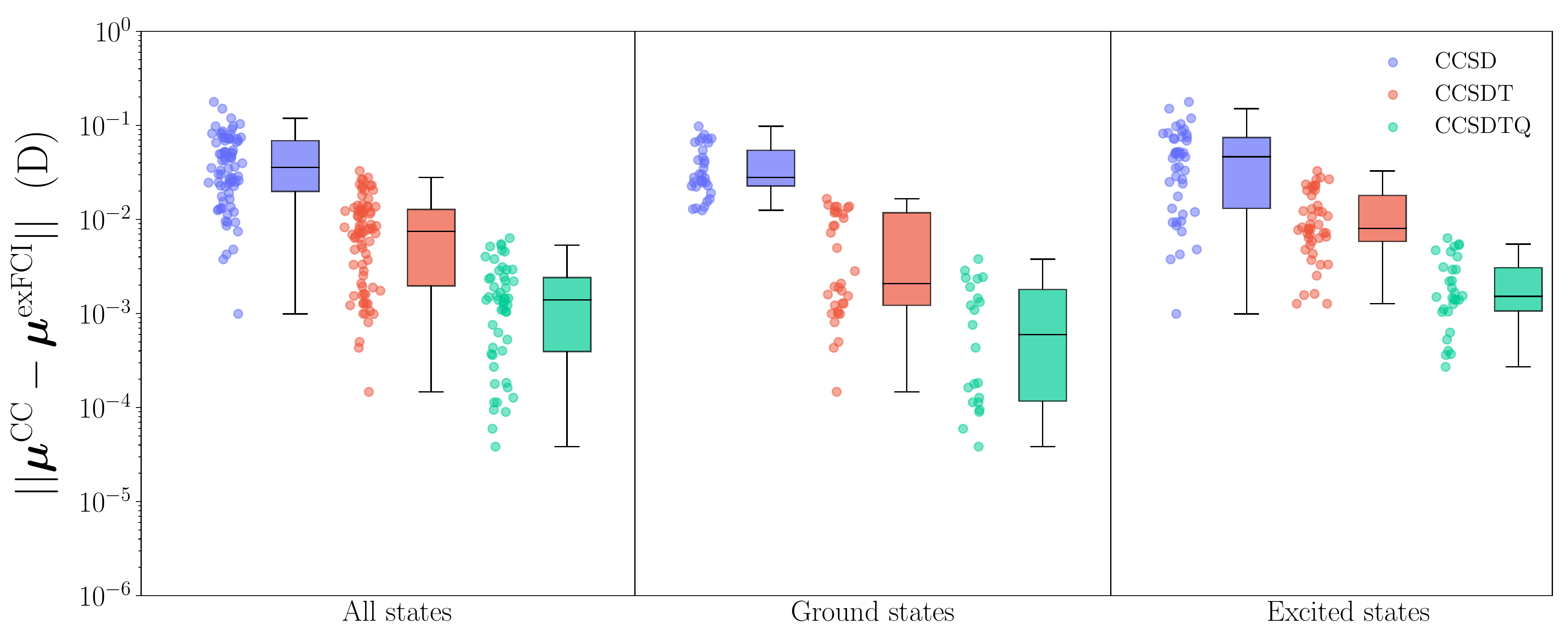}
	\caption{Box plots of the error in the ground- and excited-state dipole moments (with respect to exFCI) obtained with the CCSD (blue), CCSDT (red), and CCSDTQ (green) levels for all basis sets listed in the {\SupMat}.}
	\label{fig:dip_err_cc}
\end{figure*}

\begin{table*}
	\caption{Statistical measures associated with the errors (with respect to exFCI) of ground-state (GS) and excited-state (ES) dipole moments computed at the CCSD, CCSDT, and CCSDTQ levels for all basis sets listed in the {\SupMat}.}
	\label{tab:stats_dip}
	\begin{ruledtabular}
		\begin{tabular}{llccccccc}
					&		&	&	\multicolumn{6}{c}{Statistical quantities (in \si{\debye})}	\\
			\cline{4-9}
			Method & State & \# states & MSE & MAE & SDE & RMSE & Max($+$) & Max($-$) \\
			\hline
			CCSD        & All     &   78 &   \num{1.3E-2} &   \num{4.5E-2} &   \num{5.5E-2} &   \num{5.7E-2} &   \num{1.5E-1} &  \num{-1.8E-1} \\
						& GS      &   33 &   \num{2.2E-2} &   \num{3.8E-2} &   \num{3.9E-2} &   \num{4.5E-2} &   \num{9.8E-2} &  \num{-6.6E-2} \\
						& ES      &   45 &   \num{6.8E-3} &   \num{5.0E-2} &   \num{6.3E-2} &   \num{6.4E-2} &   \num{1.5E-1} &  \num{-1.8E-1} \\
			\\
			CCSDT       & All     &   78 &   \num{4.5E-3} &   \num{9.1E-3} &   \num{1.1E-2} &   \num{1.2E-2} &   \num{3.3E-2} &  \num{-2.8E-2} \\
			            & GS      &   33 &   \num{3.7E-3} &   \num{5.9E-3} &   \num{7.1E-3} &   \num{8.1E-3} &   \num{1.7E-2} &  \num{-1.2E-2} \\
			            & ES      &   45 &   \num{5.2E-3} &   \num{1.1E-2} &   \num{1.3E-2} &   \num{1.4E-2} &   \num{3.3E-2} &  \num{-2.8E-2} \\
			\\
			CCSDTQ      & All     &   52 &   \num{8.9E-4} &   \num{1.8E-3} &   \num{2.2E-3} &   \num{2.4E-3} &   \num{5.3E-3} &  \num{-6.3E-3} \\
			            & GS      &   22 &   \num{4.0E-4} &   \num{1.1E-3} &   \num{1.5E-3} &   \num{1.5E-3} &   \num{3.8E-3} &  \num{-2.9E-3} \\
			            & ES      &   30 &   \num{1.3E-3} &   \num{2.3E-3} &   \num{2.6E-3} &   \num{2.9E-3} &   \num{5.3E-3} &  \num{-6.3E-3} \\
		\end{tabular}
	\end{ruledtabular}
\end{table*}

Our goal is to gauge the quality of the (orbital-relaxed) dipole moments obtained at various LR-CC levels for our set of 11 molecules, by comparing them to our near-FCI estimates.
The box plot representations of the error in ground- and excited-state dipole moments computed at the CCSD (blue), CCSDT (red), and CCSDTQ (green) levels for all basis sets listed in the {\SupMat} are represented in Fig.~\ref{fig:dip_err_cc}.
The corresponding statistical quantities are reported in Table \ref{tab:stats_dip}.
We decided not to report any trends on CCSDTQP as the error between the latter method and exFCI is of the same order of magnitude as the extrapolation errors.

Considering both the ground- and excited-state dipoles, the usual trend of systematic improvement is nicely illustrated with the MAEs going down from \SI{4.5E-2}{\debye} for CCSD to \SI{1.8E-3}{\debye} for CCSDTQ.
The inclusion of triples already provides an accuracy below \SI{E-2}{\debye} (MAE of \SI{9.1E-3}{\debye} for CCSDT), which would be classified as very accurate for most applications.
In other words, going from one excitation degree to the next one (from CCSD to CCSDT or from CCSDT to CCSDTQ) reduces most of the statistical indicators by approximately one order of magnitude.
The MSEs are positive for both ground and excited states, meaning that the magnitudes of the dipole moments tend to be overestimated by LR-CC, at least for the present set of compounds.
In addition, the largest errors are generally positive and obtained for the excited-state dipoles.
An analysis of the other statistical quantities leads to similar conclusions.
As one notices by comparing the central and right panels of Fig.~\ref{fig:dip_err_cc}, CC methods are more accurate for ground-state dipole moments than for excited-state ones, which is expected since the LR (as well as EOM) formalism is naturally biased towards the ground state.

\begin{figure*}
	\includegraphics[width=\linewidth]{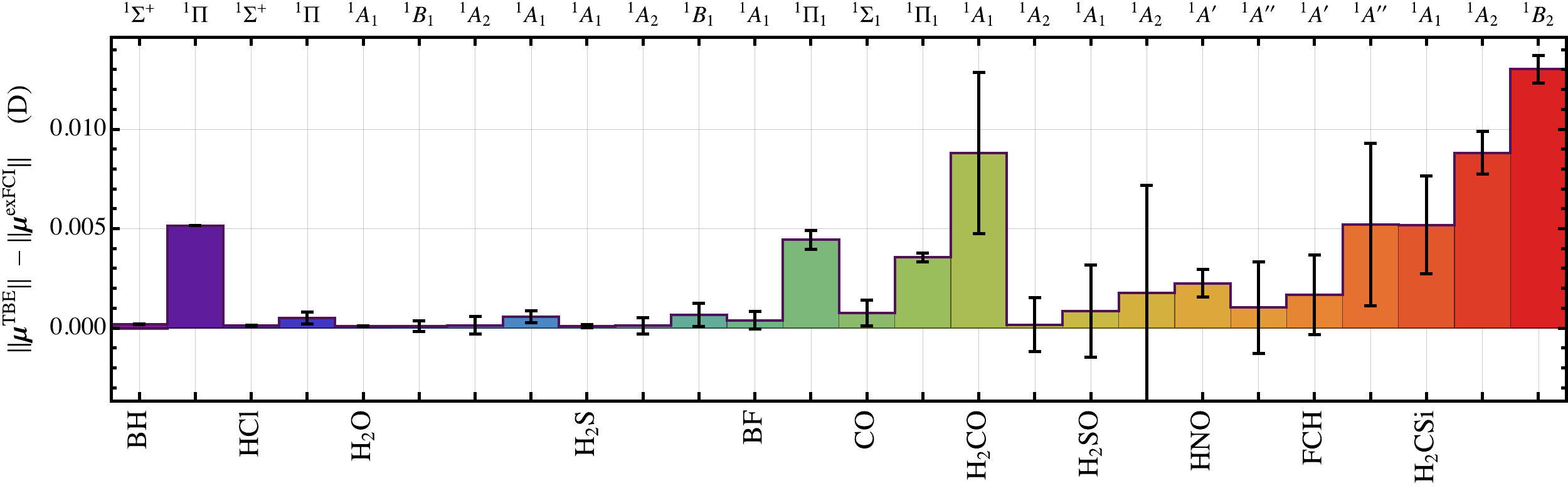}
	\caption{Comparison between the orbital-relaxed TBE dipole moments from Chrayteh \textit{et al.} \cite{Chrayteh_2021} and the present exFCI values for the 26 states considered in the present study.
	The estimated extrapolation error associated with each exFCI value is also reported.
	All these quantities have been computed in the aug-cc-pVTZ basis.}
	\label{fig:dip_tbe_err_avtz}
\end{figure*}

The exFCI/aug-cc-pVTZ values can also be compared to the orbital-relaxed TBEs obtained by Chrayteh \textit{et al.} \cite{Chrayteh_2021} for the dipole moments computed in the same basis.
The differences between these two sets of accurate data are reported in Fig.~\ref{fig:dip_tbe_err_avtz} (see Table \ref{tab:exFCI-vs-TBE} for the raw data).
Small differences are observed for \ce{BH} and \ce{BF} between the orbital-relaxed and orbital-unrelaxed dipole moments of the first excited state due to
the frozen core approximation (see Sec.~\ref{sec:intro}). \cite{Baeck_1997}
For formaldehyde, fluorocarbene, and silylidene, the convergence of the CIPSI calculations for the different states is not completely satisfactory, leading to larger uncertainties on the exFCI values, hence explaining the difference with the TBEs.
Excluding these cases, the TBEs are found to be in excellent agreement with the exFCI results with differences of few \si{\milli\debye} only.

\subsection{Oscillator strengths}

\begin{figure}
	\includegraphics[width=\linewidth]{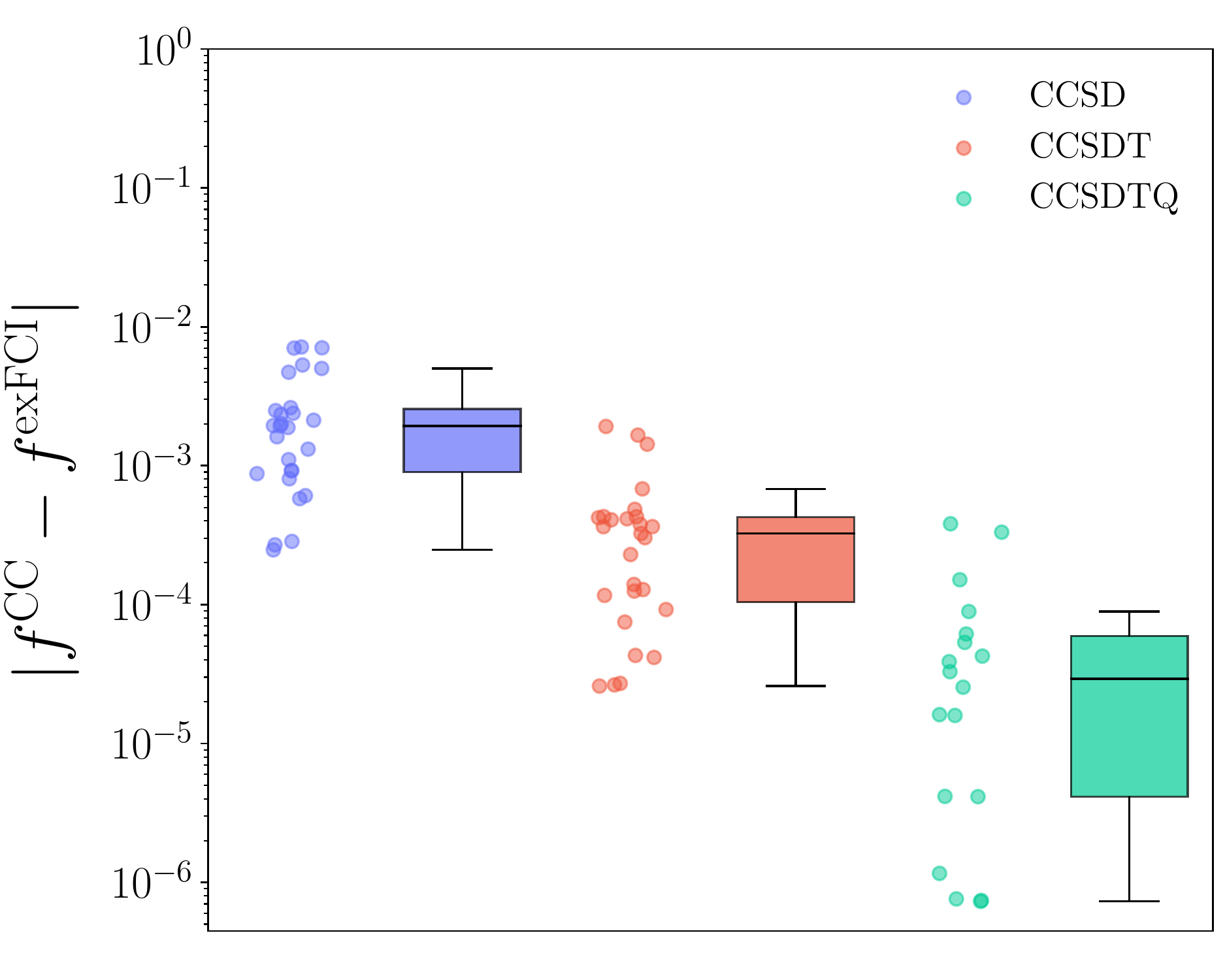}
	\caption{Box plots of the error in oscillator strengths (with respect to exFCI) computed at the CCSD (blue), CCSDT (red), and CCSDTQ (green) levels for various basis sets.}
	\label{fig:osc_err_cc}
\end{figure}

 \begin{table*}
	\caption{Statistical measures associated with the errors (with respect to exFCI) of the oscillator strengths computed in the length representation at the CCSD, CCSDT, and CCSDTQ levels for all basis sets listed in the {\SupMat}.}
	\label{tab:stats_osc}
	\begin{ruledtabular}
		\begin{tabular}{lccccccc}
            			&		&		\multicolumn{6}{c}{Statistical quantities}	\\
            			\cline{3-8}
            Method      &  \# states    & MSE        & MAE        & SDE        & RMSE       & Max($+$)  & Max($-$) \\
            \hline
            CCSD           &   27 &  \num{5.0E-4}  &   \num{2.4E-3} &   \num{3.2E-3} &   \num{3.2E-3} &   \num{7.1E-3} &  \num{-7.2E-3} \\
            CCSDT          &   27 &  \num{-1.6E-4} &   \num{4.1E-4} &   \num{6.1E-4} &   \num{6.3E-4} &   \num{6.8E-4} &  \num{-1.9E-3} \\
            CCSDTQ         &   18 &  \num{-4.4E-5} &   \num{7.0E-5} &   \num{1.2E-4} &   \num{1.3E-4} &   \num{8.9E-5} &  \num{-3.8E-4} \\
		\end{tabular}
	\end{ruledtabular}
\end{table*}

Let us now focus on the performance of CC methods for oscillator strengths by comparing them to exFCI.
The corresponding statistical analysis considering length representation (for all basis sets listed in the {\SupMat}) can be found in Table \ref{tab:stats_osc}.
The box plots of the errors associated with CCSD, CCSDT, and CCSDTQ are represented in Fig.~\ref{fig:osc_err_cc}.

Concerning the statistics, the results gathered in Table \ref{tab:stats_osc} show that the MSEs of the different CC methods are close to zero, sometimes positive and sometimes negative, meaning that one cannot conclude if the oscillator strengths tend to be overestimated or underestimated.
Also, similarly to the dipole moments, going from one excitation degree to the next one reduces the error and all the statistical quantities by approximately one order of magnitude (see Fig.~\ref{fig:osc_err_cc}).
Overall, we have found that the oscillator strengths are easier to converge at the SCI level than the individual dipole moments.
We note that CCSDT  provides a MAE well below \num{E-3}, which is sufficient for most applications.

Table \ref{tab:exFCI-vs-TBE} reports the oscillator strengths at the exFCI/aug-cc-pVTZ level in the different representations for the 9 dipole-allowed transitions considered in the present study.
The corresponding TBEs extracted from the work of Chrayteh \textit{et al}. \cite{Chrayteh_2021} and computed in the length representation are also listed for comparison purposes.
As one would see, there is a perfect agreement between the two sets of data, which confirms the quality of the TBEs reported in Ref.~\onlinecite{Chrayteh_2021}.
In Table \ref{tab:exFCI-vs-TBE}, we also report the oscillator strengths computed in the velocity and mixed representations.
Except for a few valence transitions, they do not significantly differ from their length counterparts.
In each case, $f^\text{LV}$ can be fairly well approximated by the averaged value of $f^\text{L}$ and $f^\text{V}$, as expected from their mathematical definitions (see Sec.~\ref{sec:expval}).

\section{Concluding Remarks}
\label{sec:ccl}

In this work, we have implemented the computation of the ground- and excited-state dipole moments, as well as the oscillator strengths, at the SCI level using the expectation value formalism.
Thanks to an efficient implementation of the SCI+PT2 method known as CIPSI and tailored extrapolation procedures, we have been able to reach near-FCI accuracy for these properties in the case of 11 small molecules.
In most cases, the magnitude of the dipole moments was computed with an accuracy of few \si{\milli\debye}.
Similarly, we have reached an accuracy of the order of \num{E-4}{} for the oscillator strengths in the length, velocity, and mixed representations.
Of course, the accuracy is constrained by the size of the Hilbert space, and reaching such a level of precision is hence limited to compact systems.
Nevertheless, the principal ambitions of the present work are (i) to illustrate how one can reach near-FCI quality for electronic properties with SCI+PT2 methods, and (ii) how they can be useful to estimate errors in state-of-the-art CC models which are usually challenging to assess due to the lack of reference data.
The main highlights of the present benchmark are that CCSDT is accurate enough for most practical applications, while CCSD produces MAEs of \SI{3.8E-2}{\debye} (\SI{5.0E-2}{\debye}) and \num{2.4E-3} for ground-state (excited-state) dipole moments and oscillator strengths, respectively.

As a perspective, the present strategy could be further improved by taking into account the (perturbative) first-order wave function in the computation of the expectation values.
This would be particularly useful to tackle larger systems.
Work along these lines is currently in progress in our group.

\section*{Supporting Information}
\label{sec:supmat}
See the {\SupMat} for the raw data associated with each figure and table, the molecular geometries, and the Hartree-Fock energies corresponding to the different basis sets.
The dipole moments and oscillator strengths of the 11 molecules are also reported for all levels of theory and basis sets considered in the manuscript.

\acknowledgements{
This project has received funding from the European Research Council (ERC) under the European Union's Horizon 2020 research and innovation programme (Grant agreement No.~863481).
This work used the HPC resources from CALMIP (Toulouse) under allocation 2022-18005 and from the CCIPL center (Nantes).}


\bibliography{dipole_cipsi}

\end{document}